\documentclass{ws-jtcc}

\def \bc{\begin{center}}
\def \ec{\end{center}}
\def \be{\begin{equation}}
\def \ee{\end{equation}}
\def \bt{\begin{tabular}}
\def \et{\end{tabular}}
\def \ba{\begin{array*}}
\def \ea{\end{array*}}
\def \hs1{\hspace*{1cm}}

\def \bfr{\begin{flushright}}
\def \efr{\end{flushright}}
\def \bfl{\begin{flushleft}}
\def \efl{\end{flushleft}}
\def \bea{\begin{eqnarray*}}
\def \eea{\end{eqnarray*}}

\newcommand{\krn}{\kern -2pt}

\begin{document}

\markboth{Geetha Gopakumar et al}
{RPA effects in allowed and PNC induced E1 transitions}

%
\catchline{}{}{}{}{}
%

\title{Random Phase Approximation For Allowed and Parity Non-conserving
Electric Dipole Transition Amplitudes and its Connection with Many-
Body Perturbation Theory  and Coupled-Cluster Theory}

\author{GEETHA GOPAKUMAR}
\address{Department of Applied Chemistry, University of Tokyo, Japan.\\
geetha@qcl.t.u-tokyo.ac.jp\\
http://www.u-tokyo.ac.jp}
\author{CHIRANJIB SUR, BHANU PRATAP DAS, RAJAT K. CHAUDHURI}
\address{Indian Institute of Astrophysics, Bangalore, India\\
csur@iiap.res.in, das@iiap.res.in, rajat@iiap.res.in}
\author{DEBASHIS MUKHERJEE}
\address{Indian Association for the Cultivation of Science,
Kolkata, India\\
pcdm@iacs.res.in}
\author{KIMIHIKO HIRAO}
\address{Department of Applied Chemistry, University of Tokyo,
Japan\\
hirao@qcl.t.u-tokyo.ac.jp}
\maketitle
\begin{history}
\received{Day Month Year}
\revised{Day Month Year}
\accepted{Day Month Year}
\end{history}

\begin{abstract}
The connections between the Random Phase Approximation (RPA) and Many-Body
Perturbation Theory (MBPT) and its all order generalisation, the Coupled-
Cluster Theory (CCT) have been explored. Explicit expressions have been
derived for the electric dipole amplitudes for allowed and forbidden
transitions induced by the parity non-conserving neutral weak interaction.
The Goldstone diagrams associated with the RPA terms in both cases are 
shown to arise in MBPT and CCT and the numerical verification 
of this relationship is made for the allowed electric dipole transitions. 

\end{abstract}


\section{INTRODUCTION}

The Random Phase Approximation (RPA) has been successfully used in
calculating core polarisation/relaxation effects in a variety of
properties \cite{bloch}. It is an approximate many-body theory that has been
formulated in a number of different but equivalent ways \cite{dalgarno}. 
However, its connection with MBPT and CCT has not been explored in detail
to the best of our knowledge. We focus on this particular point
in this paper.

The outline of the paper is as follows. We first derive an effective
operator in MBPT for electric dipole ($E1$) transition amplitudes for allowed 
and forbidden transitions induced by the Parity Non Conserving (PNC) neutral 
current weak interaction \cite{commins}. From the general expression in MBPT, 
we consider diagrammatically all the RPA diagrams pertaining to zeroth and 
first order in the residual Coulomb interaction.
This is followed by theoretical understanding of these effects starting from 
Hartree Fock (HF) equations to show how these effects can be represented as 
linear equations which in turn can be solved to self-consistency taking the
residual Coulomb interaction to all orders. 
At the end, starting from the basic formalism for CC method, we compare the 
RPA diagrams both theoretically and numerically with the corresponding 
diagrams in CCT.

\section{General form of an effective operator in MBPT}

In perturbation theory, the functional space for the wave function is separated into two parts; a model
space (P) and an orthogonal space (Q). The basic idea of such a division is to find an effective operator which act
only within the limited model space but which generate the same result as do the original operators acting on 
the entire functional space. Here in the sections which follow, we derive an effective operator for allowed
and PNC induced E1 transitions using perturbation theory. 

\subsection{General form of an effective operator for allowed E1 transitions}

\begin{figure}[th]
\centerline{\psfig{file=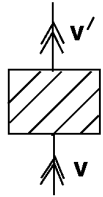,width=0.5in}}
\caption{Form of the diagram for $E1$ with $v$ and $v^{\prime}$ denoting valence lines.}
\label{form}
\end{figure}

\begin{figure}[th]
\centerline{\psfig{file=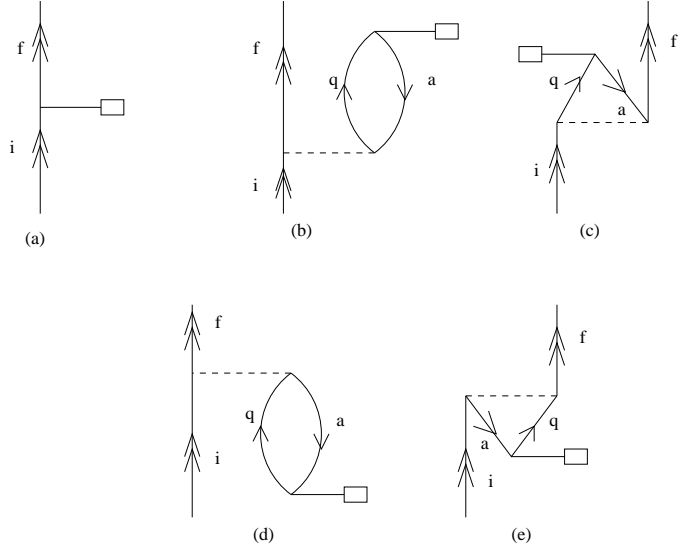,width=3.5in}}
\caption{Diagrams representing $E1^{(1)}$ contributions with diagrams (2b) and 
(2c) denoting the 
direct and exchange contributions with diagrams (2d) and (2e) with their 
Hermitian conjugate part. Here
square sign represents Dipole and the dotted lines denotes Coulomb operator. 
Fig (2a) corresponds to the Dirac-Fock part.}
\label{e11}
\end{figure}

We start the derivation with the total Hamiltonian as
\be
H =H_{0} +V_{es}
\ee
where
$H_{0}$ and $V_{es}$ are the unperturbed one-electron and perturbed two-electron operators. 
If $|\Psi_{0}\rangle$ is the atomic state function (ASF), then it satisfies the equation, 
\be
H_{0}|\Psi_{0}\rangle=E_{0}|\Psi_{0}\rangle.
\ee
where $E_{0}$ is the energy eigenvalue of the ASF. 
Here we are interested in allowed $E1$ dipole transitions between the eigen states of the atomic 
Hamiltonian $H$ as given by 
\be
E1 = \langle{\Psi}_{\beta}|D|{\Psi}_{\alpha}\rangle
\ee
where $\alpha$ and $\beta$ denotes two different ASFs of different parity and $D$ the dipole operator.
Considering $\Omega$ as the wave operator which upon acting on an
unperturbed part generates the exact state, the observable $E1$ reduces to
\be
E1 = \langle \Psi_{\beta}^{(0)}|\Omega^{\prime\prime}D\Omega^{\prime}
|\Psi_{\alpha}^{(0)}\rangle 
\label{e4144}
\ee
Here, $\Psi_{\alpha}$ and $\Psi_{\beta}$ denotes the initial and final atomic state functions.
Once $\Omega$ is known, $E1$ can be computed. Starting from Bloch equation\cite{bloch} and
considering only terms of different orders in Coulomb interactions, we get $E1$ with 
$D_{eff}$ operator defined as
\be
{D_{eff}} = \sum_{m=0}^{\infty} [\Omega_{es}^{\prime\prime (m)}]{}^{\dagger} D 
\sum_{n=0}^{\infty} [\Omega_{es}^{\prime
(n)}] 
\ee
where $\Omega_{es}^{m,n}$ refers to the  wave operator with different orders 
of the residual Coulomb interaction denoted by $m$ and $n$ 
corresponding to the initial and final states. We consider $D_{eff}$ to have only connected diagrams. 
By putting $n=0$ and $m=0$, it can be
verified that it reduces to unperturbed  contribution. Hence, a very general effective operator can be
rewritten as
\be
D_{eff}^{(n)}= \sum_{m=0}^{n}[\Omega_{es}^{\prime\prime(m)}
]{}^{\dagger} D [\Omega_{es}^{\prime(n-m)}].
\ee
Hence $E1$ to any order $n$ is given by
\be
E1^{(n)}= \langle\Psi_{\beta}^{(0)} |D_{eff}^{(n)}|\Psi_{\alpha}^{(0)}\rangle.
\ee
For $E1^{(1)}$ with one order in Coulomb interaction, $D_{eff}^{(1)}$ reduces to
\be
D_{eff}^{(1)} = D R^{\prime} V_{es}\hat{P} 
+(R^{\prime\prime}V_{es}\hat{P}){}^{\dagger}D
\ee
where $R^{\prime}$ and $R^{\prime\prime}$ are the resolvent operators given by
\begin{eqnarray}
R^{\prime} = \sum_{\gamma \notin M} \frac {|\Phi_{\gamma}\rangle\langle
\Phi_{\gamma}|}
{E_{\alpha} - E_{0}^{\gamma}} \nonumber \\
R^{\prime\prime} = \sum_{\gamma \notin M} \frac {|\Phi_{\gamma}\rangle\langle
\Phi_{\gamma}|}
{E_{\beta} - E_{0}^{\gamma}}.
\end{eqnarray}
Here  $\alpha$, $\beta$ are in the model (P) and $\gamma$ in the orthogonal (Q) space with
$E$ defining their corresponding eigenvalues and $M$ being the dimension of the model space considered. 
Considering the ASFs ($\alpha, \beta$) to be single determinant with single open valence shell 
we consider diagrams only of the form given in Fig. \ref{form}.
The zeroth order $E1$ diagrams are represented in Fig. \ref{e11}. Here diagrams 
(2b) and (2c) denote the direct and the exchange parts respectively and 
diagrams (2d) and (2e) their Hermitian conjugate parts. Higher order 
$E1$ diagrams can be obtained by taking different orders of $D_{eff}^{(n)}$ terms and contracting
by Wick's theorem to obtain diagrams only of the form given in Fig. \ref{form}.
In the next section, we follow similar derivation by taking $PNC$ as an additional
perturbation along with Coulomb operator. 

\subsection{General form of effective operator for PNC induced allowed $E1$ transitions}

We start the derivation with the total Hamiltonian as
\be
H =H_{0} +V_{es}+ H_{PNC}
\ee
where
$V_{es}$ and $H_{PNC}$ are two-electron and one-electron operators and 
$H_{PNC}$ is expressed as
\be
H_{PNC}=\frac{G_{F}}{2\sqrt{2}}Q_{W}\sum_{e}\gamma_{5}^{e}\rho(r_{e}),\label{pnc-3}
\ee
with 
\be
Q_{W}=2\left[ZC_{1p}+NC_{1n}\right].\label{pnc-4}
\ee
 Here $Z$ and $N$ are the number of protons and neutrons respectively
and $C_{1p}$ and $C_{1n}$ are the vector (nucleon) - axial vector
(electron) coupling coefficients whereas $G_{F}$ is the Fermi coupling
constant and $\rho(r_{e})$ is the normalised nucleon number density.
The matrix element of $H_{PNC}$ scales as $Z^{3}$. 
Treating $H_{PNC}$ as a first order perturbation, an ASF can be written 
as a state of mixed parity.
\be
|\tilde{\Psi}\rangle = |\Psi^{(0)}\rangle + |\Psi^{}(corre)\rangle
\ee
where $|\Psi^{(0)}\rangle$ denotes the unperturbed part which is
even/odd under parity and
$|\Psi^{}(corre)\rangle$ denotes the correction due to the perturbation
which is opposite in parity with respect to the unperturbed part.
Due to the mixing of parity in the ASFs, one can expect a non zero
electric dipole transition amplitude between states of same parity
denoted by $E1PNC$ as given by
\be
E1PNC = \langle\tilde{\Psi}_{\beta}|D|\tilde{\Psi}_{\alpha}\rangle
\ee
where $\alpha$ and $\beta$ denotes two different ASFs.
Let us consider that we have only one parity(even$+$/odd$-$) in the model space. 
Considering $\Omega$ as the wave operator which upon acting on an
unperturbed part generates the exact state, the observable $E1PNC$ reduces to
\be
E1PNC = \langle \Psi_{\beta}^{+(0)}|\Omega^{+ \prime\prime}D\Omega^{\prime}
|\Psi_{\alpha}^{+(0)}\rangle \label{e41441}
\ee
where the single and double prime on $\Omega$ denotes the perturbation on the initial
and final states. 
Once $\Omega$ is known, $E1PNC$ can be computed.
Starting from Bloch\cite{bloch} equation, a very general effective operator 
for various orders of residual Coulomb interaction with one order in $H_{PNC}$ can be derived similar to
the previous case. Defining $\Omega_{es}$ with various orders of residual Coulomb interaction
and $\Omega_{PNC}$ with one order in $H_{PNC}$ and different orders of residual Coulomb interaction, we can derive
the general effective operator as
\begin{eqnarray}
D_{eff}^{(n)}= \sum_{m=0}^{n}[\Omega_{es}^{\prime\prime(m)}
+\Omega_{PNC}^{\prime\prime (m)}]^{\dagger} D [\Omega_{es}^{\prime(n-m)}+
\Omega_{PNC}^{\prime (n-m)}].
\end{eqnarray}
where $m$ and $n$ refers to various orders of perturbation on initial and final states. 
For $n=1$ with one order in $H_{PNC}$ and zero orders of residual Coulomb interaction,
$D_{eff}^{(1)}$ reduces to 
\bea
D_{eff}^{(1)} = D R^{\prime}H_{PNC}\hat{P} + (R^{\prime\prime}H_{PNC}\hat{P})^{\dagger}D
\eea
where $R$ and $R^{\prime}$ are defined as in the above case. By considering the ASFs to be single 
determinant with single open valence line and the fact that $D$ and
$H_{PNC}$ are single particle operators, we consider diagrams only of the form given by 
Fig.\ref{form}. 
All the possible zeroth and first order diagrams with the corresponding 
expressions are given in reference\cite{geethacphf}. Out of that we are interested only in the 
RPA kind of diagrams which can be represented as linear equation starting from Hartree Fock (HF) equation and solved to all orders
of residual Coulomb interactions.

\section{Core Polarisation effects (RPA) in MBPT}
Core polarisation effects arise from the residual Coulomb interaction which is
treated as a perturbation in MBPT. At each order of perturbation there is a 
single excitation from the core. The remaining correlation effects involving 
multiple excitations will not be discussed in this work. 
In RPA\cite{mar,san,dzuba} theory, the core electrons get perturbed in the 
presence of an oscillating electric field. 
By taking these orbital modifications into account in DF potential leads to 
coupled equations for the electric dipole perturbed functions. In the 
subsections below, we show the equivalent terms/diagrams for the above effects 
in MBPT.

\subsection{RPA effects in MBPT for allowed $E1$ transitions}
\begin{figure}[th]
\centerline{\psfig{file=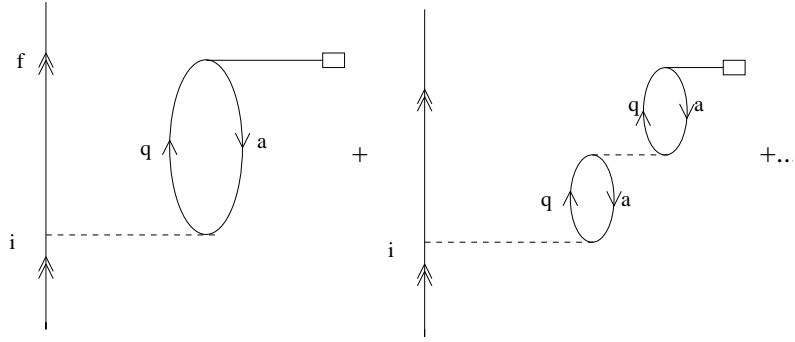,width=4.1in}}
\caption{The kind of RPA diagrams which are solved to all orders by representing the effects as 
linear equation and solving to self-consistency. Exchange and Hermitian conjugate diagrams are also 
included.}
\label{rpa_allorder}
\end{figure}

The $E1$ transition amplitude has been given earlier in Eq.(3). This transition
amplitude in the zeroth and first order can be expressed as 
\be
E1^{(0)}=\langle \Psi^{(0)}_{\beta}|D|\Psi^{(0)}_{\alpha}\rangle
\ee
and
\be
E1^{(1)}=\langle \Psi^{(0)}_{\beta}|D|\Psi^{(1)}_{\alpha}\rangle
        +\langle \Psi^{(1)}_{\beta}|D|\Psi^{(0)}_{\alpha}\rangle
\ee
respectively.
Using the general diagrammatic rules, all the possible diagrams of the form given in Fig. \ref{form} can be obtained. This is
given in Fig. \ref{e11}. Converting the diagrams to expressions, we get
\be
E1^{(0)}_{fi} = \langle f|D|i\rangle
\ee
and
\be
E^{(1)}_{fi} = \sum_{aq} \frac {\langle f q|\tilde{V}_{es}|i a\rangle \langle a|D|q\rangle} {\epsilon_{a} - \epsilon_{q} +\epsilon_{i} -
\epsilon_{f}} + 
\sum_{aq} \frac {\langle f a|\tilde{V}_{es}|i q\rangle \langle q|D|a\rangle} {\epsilon_{a} - \epsilon_{q} -\epsilon_{i} +
\epsilon_{f}} 
\ee
Here, we define $a$ and $q$ to be core and virtual orbitals with their corresponding single particle orbital energies denoted by 
$\epsilon$. Tilde on $V_{es}$ refers to the inclusion of exchange terms. From the above expression, it is clear that due to
residual Coulomb interaction at first order the core orbital `$a$' is excited to a virtual orbitals defined as `$q$'. 
Hence the above expression denotes the first order RPA  contribution in MBPT for allowed $E1$ transitions.
Defining $\omega = \epsilon_{f}-\epsilon_{i}$, the higher
order RPA diagrams can be obtained by solving the recursive relation as given by
\be
E^{(n)}_{fi} = \sum_{aq} \frac {\langle f q|\tilde{V}_{es}|i a\rangle 
E^{(n-1)}_{aq}}{\epsilon_{a} - \epsilon_{q} -\omega} + 
\sum_{aq} \frac {\langle f a|\tilde{V}_{es}|i q\rangle 
E^{(n-1)}_{qa}}{\epsilon_{a} - \epsilon_{q} + \omega} 
\ee
where $n$ denotes the order of residual Coulomb interaction. Determining the above 
equation to self consistency 
first for core to virtual amplitudes and then using it for valence to virtual 
amplitudes is equivalent to taking all the RPA diagrams of the kind 
represented in Fig. \ref{rpa_allorder} to all orders. 

\subsection{RPA effects in MBPT for PNC induced $E1$ transitions}
\begin{figure}[th]
\centerline{\psfig{file=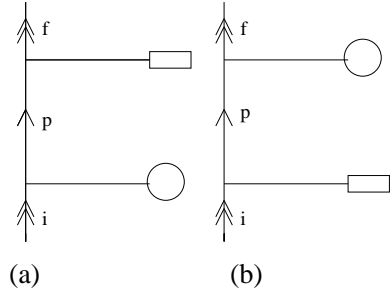,width=2.0in}}
\caption{Diagrams representing $E1PNC^{(1)}$ expression for RPA effects. The  
square and circle denotes dipole and PNC respectively.}
\label{rpa1_lower}
\end{figure}

The influence by an external oscillating electric field on the single particle orbitals of an 
atom can be obtained by solving DF equation in that field. 
By taking the matrix elements of th PNC operator between these perturbed states gives rise
to the electric dipole transition amplitude which we are interested in. 
In the lowest order, we get 
\be
E1PNC^{(1)}= \langle f^{D}|H_{PNC}|i\rangle + \langle f|H_{PNC}|i^{D}\rangle
\ee
where we have defined
\be
|(i,f)^{D}\rangle = \sum_{I} \frac{|I\rangle\langle I|D|(i,f)\rangle}
{(\epsilon_{(i,f)} -\epsilon_I)} \label{rpaa444}
\ee
Here $\epsilon_{I}$ denotes the intermediate single particle energy. With additional 
mathematical manipulations (as given in reference\cite{geethacphf})
one can derive the all order RPA equation which in turn can also 
be represented as a linear equation and solved to all orders.
The lowest order RPA contribution is shown in Fig. \ref{rpa1_lower}.
Similarly by taking one order in residual Coulomb interaction,
the second order electric dipole transition amplitude takes the form
\begin{eqnarray}
\label{111444}
E1PNC^{(2)} &=&\sum_{p} \frac {\langle f|H_{PNC}|p\rangle\langle p|D|i\rangle}
{(\epsilon_i+\omega-\epsilon_p)} \\ \nonumber
& + &\sum_{paq} \frac {\langle f|H_{PNC}|p\rangle\langle pa|\tilde{V}_{es}|iq\rangle\langle
q|D|a\rangle}{(\epsilon_i+\omega-\epsilon_p)(\epsilon_a+\omega-\epsilon_q)} \\
\nonumber
& - & \sum_{paq} \frac {\langle f|H_{PNC}|p\rangle\langle pq|\tilde{V}_{es}|ia\rangle\langle
a|D|q\rangle}{(\epsilon_i+\omega-\epsilon_p)(\epsilon_q+\omega-\epsilon_a)}
\end{eqnarray}
with `p,q' and `a' refers to virtual and core orbitals. The above
equation describes the first order RPA effect as it involves the excitation of 
the core orbital `a' to a virtual orbital `q' 
through the residual Coulomb interaction. The difference lies in the presence of the $H_{PNC}$ perturbation acting either on the initial
or final states.
\begin{figure}[th]
\centerline{\psfig{file=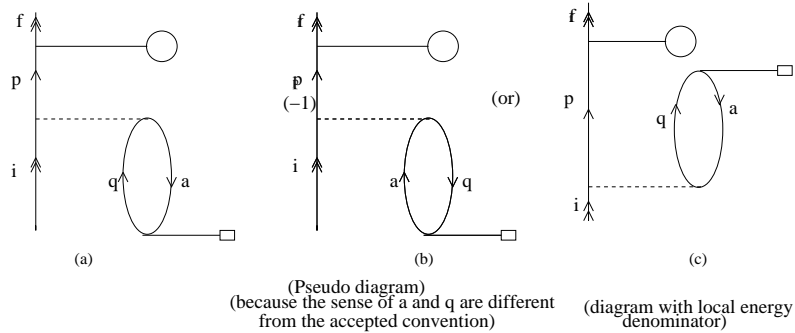,width=4.1in}}
\caption{Diagrams representing $E1PNC^{(2)}$ expression for RPA effects with square and circle sign representing
dipole and PNC perturbations. The dotted line denotes Coulomb interaction.}
\label{rpa1}
\end{figure}
Comparing diagrammatically, the first and second terms of Eq. (\ref{111444})
is equivalent to 
the RPA diagram in Fig. (4a) and (5a) respectively. The third term is 
represented in two different ways as RPA diagrams in Figs.(5b) and (5c). 
In the first case, in Fig. (5b),
the sense of the core and particle is different from the accepted convention with respect to D 
vertex and hence it is called a pseudo diagram. Whereas in the second case, by taking the negative
sign inside the expression for $E1PNC^{(2)}$, Fig. (5c) can be interpreted by
local energy denominator for both D and 
Coulomb vertex. 

Comparing the above terms with all the first order MBPT terms as discussed in \cite{geethacphf}, we can 
find that the pseudo/local energy denominator diagram can be obtained by adding two MBPT diagrams 
as shown in Figs. (6b) and (6c) respectively.  Similarly, the Hermitian 
conjugate pseudo diagram can also be obtained by adding the 
corresponding Hermitian conjugate MBPT diagrams. Fig. (6a) represents the
normal RPA diagram as in the Fig. (5a). 

\begin{figure}[th]
\centerline{\psfig{file=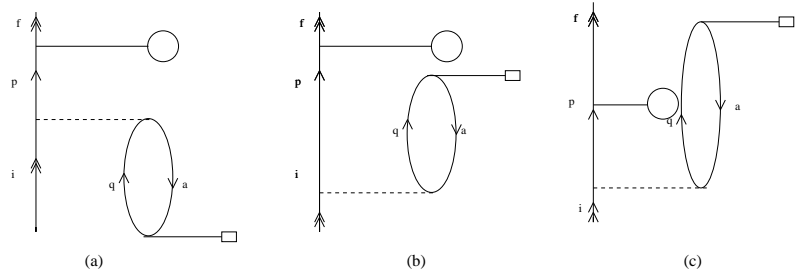,width=4.1in}}
\caption{MBPT diagrams corresponding to RPA. MBPT(1) corresponds to the normal RPA diagram. 
MBPT(2) and MBPT(3) are added to get the second term of the $E1PNC^{(2)}$ expression for RPA 
or in other words the pseudo/local energy denominator diagram.}
\label{rpa2}
\end{figure}
Similar to the allowed $E1$ transition, in the case of PNC, the influence of
$H_{PNC}$ as a perturbation can be treated in the framework of RPA.  
In the next section, we show the basic formulation for CC method and show
how RPA terms arises in the above mentioned method. 

\section{RPA effects in Coupled Cluster (CC) method}
We have so far been discussing about perturbation theory in which some of the 
terms of the perturbation theory were grouped together and evaluated to all 
orders. 
This is similar in spirit to the Coupled Cluster method where a particular 
class of MBPT diagrams is calculated to all orders in the residual Coulomb 
interaction.

The many-electron wave function in CC method is given by
\be
|\Psi_{0}\rangle=e^{(T)}\{e^{S}\}|\Phi_{0}\rangle,
\ee
where $|\Phi_{0}\rangle$ is the reference state and $T$ and $S$ are the cluster operators
which considers excitations from core and valence to virtual orbitals. 
\begin{figure}[th]
\centerline{\psfig{file=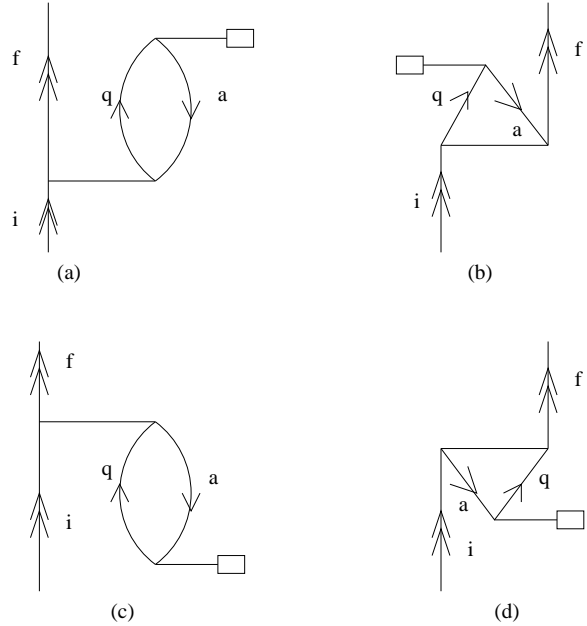,width=3.0in}}
\caption{Coupled cluster counter part of fig 2. Here the dotted line is 
replaced by a solid line representing an all-order Coulomb interaction vertex. }
\label{cp-diag}
\end{figure}

The Hamiltonian in the presence of the PNC weak interaction is  given by
\be
H=H_{a} +  H_{PNC}
\ee
where $H_{a}$ and $H_{PNC}$ denotes the atomic Hamiltonian and the PNC perturbation with $G_{F}$ showing the 
strength of the perturbation.  Hence 
\bea
T &=& T^{(0)} + G_{F} T^{(1)}, \\ \nonumber
S &=& S^{(0)} + G_{F} S^{(1)}. \nonumber
\eea
Here we denote the $T$ and $S$ with a superscript `0,1' as unperturbed and 
as PNC perturbed cluster amplitudes. Once the cluster amplitudes are computed, 
$E1PNC$ can be obtained by taking
one order in $H_{PNC}$ perturbation in initial/final states either through $T$ or $S$ cluster 
amplitudes \cite{thesis}
connected by the $D$ operator.
Whereas for the allowed $E1$ calculations, the unperturbed cluster amplitudes 
in the initial/final states are connected by the $D$ operator.
In the sections below, the RPA diagrams are compared both diagrammatically and numerically with CC for allowed $E1$ transitions.

\subsection{Diagrammatic comparison of RPA diagrams corresponding to CC 
              diagrams for allowed and PNC induced $E1$ transitions}
We now consider CC diagrams corresponding to the RPA effects.
In CC theory for allowed and PNC perturbed dipole 
transitions, we consider $D$ as an operator. The Coulomb operator acts as a 
perturbation for the allowed transitions, 
whereas for the PNC induced $E1$ transitions, we have PNC perturbation in 
addition to it.  
Therefore we need to first find out from which side of the dipole operator the 
perturbations act.

The diagrams for the allowed $E1$ transition amplitude in the framework of 
CC theory has been discussed in one of our earlier papers \cite{geetha-ba+}. 
The RPA diagrams given in Figs. (2b-2e) correspond to the $DS_{2}$ and 
$S^{\dagger}_{2}D$ diagrams of CCT (See Fig \ref{cp-diag}). 
In the case of PNC induced E1 transitions, the MBPT diagram (4a) and (4b) in 
Fig. \ref{rpa1_lower} corresponding
to zeroth order RPA effects can be got from $DS_{1}^{(1,0)}$ where the 
superscripts denotes the order in PNC and the residual Coulomb interaction
in the order of preference. By looking at the regular RPA diagram in
Fig. (\ref{rpa1}a) contributing to first order in Coulomb and PNC, we find that it is equivalent to  
$DS_{2}^{(1,1)}$ diagram and its Hermitian conjugate in CC theory. Whereas the pseudo/local energy denominator 
diagram, which is shown as equivalent to addition of MBPT diagrams in 
Fig. (\ref{rpa2}b) and (\ref{rpa2}c) can
be got from the term $S_{1}^{\dagger(1,0)}DS_{2}^{(1,0)}$ and its Hermitian conjugate respectively. 

By comparing the terms in CC method for allowed and PNC induced E1 transitions,
we find that the additional diagrams arises solely due to the presence of PNC 
operator acting either on the incoming or the final vertex. We find that PNC 
acting on the same vertex as Coulomb operator, leads to the regular diagram and
PNC and the residual Coulomb interaction on the initial and final vertices 
leads to the additional pseudo/local energy  diagram. 

\section{Numerical  comparison of RPA diagrams corresponding to CC diagrams for allowed $E1$ transitions}

In order to demonstrate that the RPA effects are contained in CCT we have 
compared the all order RPA results with the sum of $DS_{2}$ and 
$S^{\dagger}_{2}D$
contributions in our CC calculations for the allowed 
$3s^{2}S_{1/2}\longrightarrow3p^{2}P_{1/2}$ and 
$3s^{2}S_{1/2}\longrightarrow3p^{2}P_{3/2}$ transitions in $Na^{+}$.
We first generate the single particle DF orbitals with Na$^{+}(2p^{6})$ as the starting potential. 
The GTOs were generated using the Finite Basis Set Expansion (FBSE) method \cite{rajat} with a 
primitive basis set consisting of 35s-32p-25d-25f. With appropriate energy cutoffs in the discrete core and continuum 
virtual orbital spectrum, the calculation was done with 11s-10p-9d-8f basis. 

\begin{table}[th]
\begin{center}
\begin{tabular}{ccccc}
\hline 
&
\multicolumn{2}{c|}{$3s\longrightarrow3p_{1/2}$}&
\multicolumn{2}{c}{$3s\longrightarrow3p_{3/2}$}\tabularnewline
\hline
\hline 
&         RPA&   RPA-CC&         RPA&    RPA-CC
\tabularnewline
&            &         &            &
\tabularnewline
&       0.044&    0.044&      -0.061&    -0.062
\tabularnewline
\hline
\hline 
&            &         &            &
\tabularnewline
\end{tabular}
\end{center}

\caption{\label{rpavscp}Comparative results of $E1$ reduced matrix elements 
(in \emph {a.u.}) arising from RPA and RPA-CC ($DS_{2}+S^{\dagger}_{2}D$)
for $3s\longrightarrow3p_{1/2}$ and $3s\longrightarrow3p_{3/2}$ 
transitions in $\mathrm{Na^{+}}$.}
\end{table}

As stated earlier $DS_{2}$ and $S^{\dagger}_{2}D$ terms in CCT (Fig. 7) 
correspond to the RPA diagrams given in Figs. (2b-2e). In table \ref{rpavscp} we
give the numerical values of those two contributions for  
$3s\longrightarrow3p_{1/2}$ and $3s\longrightarrow3p_{3/2}$ transitions in
 $\mathrm{Na^{+}}$ .  
As expected the agreement between the two results is very good. 

\section{Conclusion}
Using analytical and diagrammatic techniques we have identified
the RPA effects in allowed and parity non-conserving electric
dipole transition amplitudes. We have demonstrated that these
effects arise in MBPT and CCT. Indeed the all order RPA contribution
is subsumed in CCT.
\section{Acknowledgements}
The present research was partly done at Indian Institute of Astrophysics, when the author (GG) was completing
her PhD work. This work was completed later at University of Tokyo with the support from National Institute of
Advanced Industrial Science and Technology (AIST) and later by National Research Grid Initiative (NAREGI) of 
Ministry of Education, Culture, Sports, Science and Technology (MEXT), Japan. 
One of the authors (CS) acknowledges the BRNS for project no. 2002/37/12/BRNS.

\end{document}